\documentclass[12pt,preprint]{aastex}
\usepackage{graphicx}
\usepackage{aastexug}   
\shortauthors{Rothschild \& Lingenfelter}
\shorttitle{HEXTE Observations of Cas A}

\begin{document}

\title{Limits to the Cas A $^{44}$Ti Line Flux and Constraints on the Ejecta Energy and the Compact Source}

\author{R. E. Rothschild \& R. E. Lingenfelter}
\affil{Center for Astrophysics and Space Sciences, 0424, \\
University of California at San Diego, 9500 Gilman Dr., La Jolla, CA 92093}
\email{rrothschild@ucsd.edu \& rlingenfelter@ucsd.edu}

\begin{abstract}

Two long observations of Cas A supernova remnant were made by the
\emph{Rossi X-ray Timing Explorer} in 1996 and 1997 to search for hard X-ray
line emission at 67.9 and 78.4 keV from the decay of $^{44}$Ti formed
during the supernova event. Continuum flux was detected up to 100 keV,
but the $^{44}$Ti lines were not detected. The 90\%
confidence upper limit to the line flux is 3.6$\times$10$^{-5}$ photons
cm$^{-2}$s$^{-1}$. This is consistent with the recent
\emph{BeppoSAX} detection and with the \emph{CGRO}/COMPTEL detection of
the companion transition line flux for $^{44}$Sc decay. The mean
\emph{BeppoSAX}---COMPTEL flux indicates that 1.5$\pm$0.3 $\times$10$^{-4}$M$_\odot$
of $^{44}$Ti was produced in the supernova explosion. Based upon
recent theoretical calculations, and optical observations
suggesting a WN Wolf-Rayet progenitor with an initial mass of
$\geq$25 M$_\odot$, the observed $^{44}$Ti yield implies that the
Cas A supernova ejecta energy was $\sim$2$\times$10$^{51}$ ergs,
and as a result a neutron star was formed, rather than a
black hole. We suggest Cas A is possibly in the early stages
of the AXP/SGR scenario in which the push-back
disk has yet to form, and when the disk does
form, the accretion will increase the luminosity to that of
present-day AXP/SGRs and pulsed emission will commence.

\end{abstract}


\section{Introduction}

Optical observations suggest that Cas A is the remnant of a Type
Ib/c supernova whose progenitor was a WN Wolf-Rayet star with initial
mass greater than 25 M$_\odot$ \citep{Fn,FnB,jansen}.
The discovery of a point source at the center of the remnant by
\emph{Chandra} \citep{tananbaum} is strongly indicative of the
creation of a neutron star or black hole in the event, and would
be consistent with such a progenitor.
Subsequent observations by \emph{Chandra} \citep{deepto,murray} and
\emph{XMM} \citep{mereghetti} have confirmed the point source
nature of the object, but the existence of pulsations (and therefore
a neutron star) is still questionable. Comparison of supernova
explosion models \citep{woosley,nakamura} with measurements of
the amount of $^{44}$Ti created, can be used to estimate the
remnant core mass and, therefore, the nature of the compact
object.

$^{44}$Ti is produced by explosive Si burning and the freeze out
from nuclear statistical equilibrium in supernovae, and is
believed to be the primary source of $^{44}$Ca \citep{timmes}.
$^{44}$Ti decays with a 59.2$\pm$0.6 year half-life \citep{ahmad,gorres,norman} to $^{44}$Sc producing
two nuclear lines at 67.9 and 78.4 keV of essentially equal
intensity. The $^{44}$Sc decays to the first excited state of $^{44}$Ca
which then emits a gamma-ray of 1157 keV with a 3.93 hour
half-life to reach the ground state. All three nuclear lines are
expected to have essentially the same line strength. Given the
half-life of $^{44}$Ti, the distance to Cas A (3.4 kpc;
\citet{reed}), the time since the supernova (317 yr for a mean observation date of 1997;
\citet{ashworth}), and the observed line flux, one can estimate
the amount of $^{44}$Ti created.

Attempts to detect the $^{44}$Ti lines began with the analysis of
galactic scanning observations by the germanium spectrometer on
\emph{HEAO-3} \citep{mahoney}. A 1$\sigma$ limit of 8.3$\times$10$^{-5}$photons cm$^{-2}$s$^{-1}$
from a point source anywhere in the galaxy was determined for the $^{44}$Ti
line flux. Initial results from \emph{CGRO}/COMPTEL observations of Cas
A announced the discovery of 1.157 MeV line flux from the
$^{44}$Sc decay \citep{iyudin94}, and this was revised after
further observations to be 3.3$\pm$0.6$\times$10$^{-5}$photons cm$^{-2}$s$^{-1}$
\citep{iyudin}. \emph{CGRO}/OSSE observations of Cas A provided for a
simultaneous fit to the three nuclear lines, and yielded a 99\%
confidence upper limit of 5.1$\times$10$^{-5}$photons cm$^{-2}$s$^{-1}$
\citep{the95}, with detection of the continuum radiation to 40
keV \citep{the96}. With the launches of \emph{RXTE} and \emph{BeppoSAX} in 1995
and 1996, two powerful hard X-ray instruments became available for
$^{44}$Ti line searches. The initial \emph{RXTE}/HEXTE result was a 2.4$\sigma$
detection (4.3$\pm$1.8$\times$10$^{-5}$photons cm$^{-2}$s$^{-1}$)
during one long observation and no detection (-1.4$\pm$1.7$\times$10$^{-5}$
photons cm$^{-2}$s$^{-1}$) in another
\citep{rothschild}. Similarly, initial \emph{BeppoSAX}/PDS measurements
could only claim an upper limit ($\leq$5$\times$10$^{-5}$photons
cm$^{-2}$s$^{-1}$; \citet{vink98}). After additional observations,
\citet{vink} were finally able to claim a good detection with a
flux of 2.1$\pm$0.7$\times$10$^{-5}$photons cm$^{-2}$s$^{-1}$ for the 68
and 78 keV lines from $^{44}$Ti. In this article we present a
reanalysis of the \emph{RXTE}/HEXTE observations of Cas A that were begun
before the \citet{vink} announcement of the detection.

The $>$10 keV continuum upon which the $^{44}$Ti lines lie has been fit
historically to a power law with photon index $\sim$3 or a thermal
bremsstrahlung model whose temperature is dependent upon the
extent of the high energy limit to the data. \citet{allen} used
\emph{ASCA}, \emph{RXTE}, and \emph{CGRO} data to show that the
thermal models successful below 10 keV (2 Raymond-Smith components
and a 6.4 keV Gaussian) fall well below the data above 10 keV. The
addition of a broken power law ($\Gamma$=1.8$^{+0.5}_{-0.6}$ below
15.9$^{+0.3}_{-0.4}$ keV and $\Gamma$=3.04$^{+0.15}_{-0.13}$ above)
best described the data to 120 keV. Several authors
\citep{laming,atoyan,baring,gaisser} have described the processes
that may be involved in the photon continuum production in
supernova remnants such as Cas A. A non-thermal bremsstrahlung
component may become the dominant contributor at $\sim$100 keV,
and therefore, its presence may affect the inferred photon flux
from the $^{44}$Ti lines for observations with high sensitivity
above the line complex.

\section{Observations}

\emph{RXTE} observed Cas A several times for calibration purposes (ObsIds
00022, 10418, 30804, and 40806) and twice as part of an
investigation into the $^{44}$Ti emission lines (ObsIds 10271 and
20253). The broad-band (2-60 keV) spectral results have been
published by \citet{allen}, and this analysis concentrates on the
15-240 keV data from the High Energy X-ray Timing Experiment
(HEXTE; \citet{Roth98}). The HEXTE is a
set of eight NaI(Tl)/CsI(Na) scintillation detectors grouped into two
clusters of four detectors each.
The HEXTE detectors are mechanically collimated to a
1$^\circ$ FWHM field of view, and cover the 15-250 keV range. The full
collecting area for HEXTE is 1400 cm$^2$
(spectral capability was lost from one HEXTE detector early in the
mission).

The observation dates and accumulated livetimes are given in
Table~\ref{tab:log}. Since the livetime associated with the
calibration observations is quite small compared to that of the
two long observations, the analysis presented here will be
restricted to the latter set of data.

\placetable{tab:log}

\section{Data Reduction and Analysis}

The HEXTE data were accumulated for each cluster using a script
developed at UCSD and the University of T\"{u}bingen. The cluster
data were separated into
on-source, off-source-plus and off-source-minus, since each HEXTE
cluster collects real-time background data from two independent
positions $\pm$1.5$^\circ$ to either side of the on-source
position. The two clusters' rotation axes are orthogonal to each
other, and in this way four independent background regions are
sampled. By comparing the two off-source positions for a given
cluster, one can determine if a confusing source is contaminating
one of the background regions, and if so, eliminate it from
further analysis. The background observations are
75\% as long as the on-source observations, since the time to move
the pointing position on- and off-source comes out of the
background observations. This ensures HEXTE continual on-source
coverage.

The two background histograms from each cluster were tested for a
confusing source by specifying the off-minus data as the source
and the off-plus data as the background in the HEASARC spectral
fitting program XSPEC \citep{arnaud}. The net count rate was
compared to the ideal value of 0.0, and the spectra were examined
for a continuous deviation from zero at low energies, as a sign of
another source. Table~\ref{tab:mp} gives the net 15-240 keV rates
for the two clusters for ObsIds 10271 and 20253. Both clusters in
ObsId 10271 show a negative net flux, with that for cluster A
being 2$\sigma$ below 0.0. Investigation of the net spectra of
each revealed a negative residual at 30 and 65 keV --- the
location of the prominent background lines. This is an indication
of an under- or over-estimated livetime. The HEXTE livetime
estimation is based upon a model using the Upper Level Discriminator
(250 keV) and the eXtreme Upper Level Discriminator (20 MeV)
rates, and on a daily timescale is accurate to $\sim$1\%. Changing
the livetime of cluster A off-plus by 2.1\% and cluster B off-plus
by 1.3\% brought the net rates to 0.0 within uncertainties. ObsId 20253
data required no such change in livetime.

\placetable{tab:mp}

Since the HEXTE background has emission line features at 67 keV
due to activation of the NaI(Tl) and at the lead K-lines at 74
and 85 keV due to the collimator, one must demonstrate that any
claimed line features are not due to imperfect background subtraction. The
brightest background line complex is at 30 keV, with a flux of
3.85$\times$10$^{-2}$ photons cm$^{-2}$s$^{-1}$. The 90\% upper limit to its flux in the net
Cas A spectrum is 2.18$\times$10$^{-5}$ photons cm$^{-2}$s$^{-1}$, or 0.07\% of
background. The 67 and 74 keV background lines have fluxes
of 1.9$\times$10$^{-2}$ photons cm$^{-2}$s$^{-1}$ and 1.2$\times$10$^{-2}$ photons
cm$^{-2}$s$^{-1}$, respectively. Using the percentage of background upper limit at 30 keV,
we estimate the systematic sensitivity
limit for the $^{44}$Ti lines in this observation is 1$\times$10$^{-5}$ photons
cm$^{-2}$s$^{-1}$.

Separate good time intervals were calculated for the three pointing
positions of each cluster. The good time intervals required the pointing
direction to be within 0.01$^\circ$
of the center of the Cas A remnant, its elevation above
the Earth's horizon to be greater than 10$^\circ$, and the time
since the start of the most recent South Atlantic Anomaly passage
to be greater than 20 minutes.

For each individual observation (17 each within ObsIds 10271 and
20253), the plus and minus off-source data sets were combined to
form the background data set for each cluster, and the individual
source/background data for each ObsId were summed for each cluster.
Finally, the ObsId 10271 and 20253 cluster A and B data were
summed to form a single on-source and single off-source spectral
accumulation. The resulting pulse height histogram from
the two 200 ks observations contains 226 ks livetime on-source and
170 ks of real-time background. Subsequent spectral fitting was performed using
these two files, as well as for the combined data in each ObsId.

The pulse height data were binned into single, 1 keV bins to 30 keV,
double width bins to 89, and quadruple width bins to 100 keV.
Above that energy they were combined as 100-140, 140-180, 180-220, and 220-240 keV.
Three sets of data were fitted: ObsId 10271, 20253, and their sum.
The data were then fitted with a power law
plus two gaussian lines at the expected energies for $^{44}$Ti$\rightarrow
^{44}$Sc decay (67.9 and 78.4 keV). The intensities of the two lines were linked to be
the same value, and line widths were considered: 1) narrow lines and
2) lines broadened by 2.5\% to 1.7 and 2.0 keV
respectively to account for the Cas A expansion velocity of 7500
km/s \citep{fesen}. The results are given in Table~\ref{tab:fit}.
The fit to the continuum for the two independent observations
yields consistent indices and fluxes, and the best fit to the
summed data is a photon index $\Gamma$=3.125$\pm$0.050 and
20-100 keV flux of 4.60$\pm$0.18$\times$10$^{-11}$ ergs
cm$^{-2}$s$^{-1}$keV$^{-1}$. Fitting the $^{44}$Ti line flux in
the two independent observations yielded fluxes at the -0.87$\sigma$ and 1.69$\sigma$
level, respectively. The best fit line flux to the combined data
was 1.57$\pm$2.81$\times$10$^{-5}$ photons cm$^{-2}$s$^{-1}$, which yielded a
90\% confidence upper limit on the flux from each of the
$^{44}$Ti lines of 3.6$\times$10$^{-5}$ photons cm$^{-2}$s$^{-1}$.
These values are consistent with those measured by BeppoSAX
\citep{vink}, and are unaffected by line widths from zero to 2 keV.
The continuum was detected to 100 keV, thereby
extending the maximum detected energy reported in \citet{allen}.
The data plus best fit model for the combined data are shown in Figure~\ref{fig:fit}.

\placetable{tab:fit}

Residuals to the best fit model in the case of the ObsId 10271
observation contain a large fluctuation ($\Delta \chi^2$= 8.4) in the single bin at
70 keV. Since the 2 keV width of this
bin is less than the detector resolution at that energy (10 keV),
the deviation  is considered a statistical fluctuation. This
conclusion is supported by the residuals in the ObsId 20253 data,
where the residual at that energy is significantly less ($\Delta
\chi^2$= 3.3).

In order to investigate the effect of a non-power law continuum,
we have tried two other continuum forms: 1) a broken power law
with photon index $\sim$3 up to a break energy, after which the
continuum flattens (to mimic the onset of the non-thermal
bremsstrahlung component as shown in Figure 1 of \citet{ellison}),
and 2) a high energy thermal instead of the power law to represent
a steepening of the spectrum at high energies. Fitting the 20253
data, which does not have the over-subtraction of background,
shows that a $\Gamma$=3 power law that breaks at 43 keV (best fit)
to a $\Gamma$=2.4 power law (best fit), does not present a better
fit ($\Delta \chi^2$= 0.8 for loss of 2 degrees of freedom). The
90\% upper limit on the $^{44}$Ti lines is reduced from
9.0~$\times$10$^{-5}$ photons cm$^{-2}$s$^{-1}$ to
5.8~$\times$10$^{-5}$ photons cm$^{-2}$s$^{-1}$. Using the thermal
bremsstrahlung model in place of the power law results in a worse
fit ($\Delta \chi^2$= 10.1 for the same number of degrees of freedom).
This model under-estimates the flux before and after the $^{44}$Ti
complex. The lines do become more prominent due to the dropping
continuum with the upper limit rising to 1.07~$\times$10$^{-4}$
photons cm$^{-2}$s$^{-1}$. A steepening of the continuum, as
opposed to a straight power law, is not supported by the data.

\placefigure{fig:fit}

\section{Discussion}

The observed $^{44}$Ti flux  \citep{iyudin,vink} has very interesting
implications for the mass of the progenitor, the
supernova ejecta energy, and the question of whether
the compact object left after the explosion is a
neutron star or a black hole. Taking the best estimate
of the mean COMPTEL and BeppoSAX flux in each of the $^{44}$Ti lines to be
2.7$\pm 0.5 \times 10^{-5}$ photons/cm$^2$ s, we
find that the $^{44}$Ti yield from the Cas A supernova
should be 1.5$\pm 0.3 \times 10^{-4}$ M$_{\odot}$,
assuming an age of 317 yr for a 1997 mean date of the
observations, a $^{44}$Ti half-life of 59.2 $\pm$ 0.6 yr,
and a distance of 3.4 kpc. This can be compared with theoretical
calculations of $^{44}$Ti production in core collapse supernovae.

First, we find that the $^{44}$Ti yield is roughly a factor of 2
or more greater than the expected \citep{woosley,nakamura} yields
in any core collapse supernovae with ejecta energies close to
$\sim$1$\times$10$^{51}$ ergs, which can explain the observed
properties of most (8 to 25 M$_{\odot}$) Type II supernovae.
However, the observed yield is consistent with that expected
\citep{wooslang,nakamura} from Type Ib/c supernovae of Wolf-Rayet
stars with initial masses greater than 25 M$_{\odot}$ --- implied
by the optical observations \citep{Fn,FnB,jansen}. In particular
for a pre-collapse mass of $\sim$ 4 to 6 M$_{\odot}$, that would be
expected \citep{wooslang} for a WN Wolf-Rayet star that explodes
before reaching the WC phase, the observed yield implies an ejecta
energy of $\sim$2$\times$10$^{51}$ ergs.
These calculations for WN Wolf-Rayet pre-collapse stars also all
predict Type Ib/c supernova explosions that leave a neutron star
remnant, not a black hole.

Secondly, the predicted \citep{nakamura} $^{44}$Ti/$^{56}$Ni ratio for
the Type Ib/c supernovae of such stars is $\sim 1.7\times 10^{-3}$,
which is similar to more general predictions \citep{wooshof}.
Therefore the discrepancy between the measured $^{44}$Ti yield
and the estimated brightness of the historical supernova still
seems to require the added effects \citep{Naga98} of an
asymmmetric explosion, which has also been supported by
the recent observations of \citep{Fn}.

Finally, the consistency between the observed $^{44}$Ti yield and
that calculated for a Type Ib/c supernova of a WN Wolf-Rayet
progenitor, thus would appear to rule a black hole as the point
source discovered by \emph{Chandra} \citep{tananbaum}.

Although no pulse period has yet been detected from the
Cas A compact object, comparisons of the spectral index
and luminosity of the \emph{Chandra} x-ray source with
those of other x-ray pulsars suggest \citep{deepto,mereghetti}
that it looks more like an
Anomalous X-ray Pulsar/Soft Gamma-ray Repeater (AXP/SGR)
than a radio pulsar. We point out that another indicator
for such an association is the fact that the Cas A remnant
is expanding in the warm denser phase of the interstellar
medium (e.g. \citet{higgie,ginga}), which is also the site of the bulk
of the AXP/SGRs \citep{marsden}. The majority of the radio pulsars,
on the other hand, are observed in the hot tenuous medium where
most of the core-collapse supernovae occur. Fallback disk
\citep{alpar} and push-back disk \citep{marsden} models for
AXP/SGRs utilize ejecta material to form a small accretion disk
around a conventional neutron star (i.e., with magnetic field of
10$^{10-13}$ Gauss), which provides the additional torque to
spin-down the neutron star rapidly and the accreted matter to
generate the greater X-ray luminosity seen in AXP/SGRs.
Fallback disks form within a few days from
material that cannot escape the newly formed neutron star, whereas
push-back disks form many years later from material decelerated by the
Sedov-phase reverse shock and subsequently captured by the
neutron star. We suggest Cas A is possibly in the early stages
of the AXP/SGR scenario in which the push-back
disk has yet to form, since the reverse shock has not reached the
material in the vicinity of the neutron star. Thus, one might
expect the present emission to be indicative of a conventional
cooling neutron star, as per \citet{deepto}, and when the disk does
form, the accretion will increase the luminosity to that of
present-day AXP/SGRs and pulsed emission will commence.

\acknowledgments

The authors would like to thank Jacco Vink, Shigehiro Nagataki, and
Anatoli Iyudin for fruitful discussions, and Harish Khandrika for
data processing assistance. The authors also acknowledge the useful
comments by the referee. This work was supported by NASA contract NAS5-30720.

\appendix

\newpage


\begin{deluxetable}{llcc}
\tablecaption{Log of Observations\label{tab:log}}
\tablewidth{0pt}
\tablehead{
\colhead{ObsId} & \colhead{Dates of Observations} & \colhead{On-Source Livetime} & \colhead{Background
Livetime}\\
 & & (ks) & (ks)
}
\startdata
10271 & 20-28 April 1997 & 120.0  & 90.0 \\
20253 & 31 March - 17 April 1996 & 106.2  & 79.8 \\
00022 & 20 January 1996 & 12.3  & 10.8  \\
10418 & 1-2 August 1996 & 2.7  & 2.0  \\
30804 & 10 March 1998 & 2.9 & 1.5  \\
40806 & 23-25 March 1999 & 6.8  & 5.8  \\
      & 5 August 1999 & & \\
\enddata
\end{deluxetable}

\clearpage

\begin{deluxetable}{lll}
\tablecaption{Comparison of Off-Source Observations\label{tab:mp}}
\tablewidth{0pt}
\tablehead{
\colhead{ObsId} & \colhead{Cluster}  & \colhead{Net Count Rate}\\
 & & (c/s)
}
\startdata
10271 & A & -0.266$\pm$0.109 \\
      & B & -0.111$\pm$0.091 \\
20253 & A & -0.056$\pm$0.116 \\
      & B & -0.031$\pm$0.102 \\
\enddata
\end{deluxetable}

\clearpage

\begin{deluxetable}{llll}
\tablecaption{Best Fit Cas A Spectral Parameters\label{tab:fit}}
\tablewidth{0pt}
\tablehead{
\colhead{Parameter} & \colhead{ObsId 10271} & \colhead{ObsId 20253} & \colhead{Combined}
}
\startdata
Photon Index & 3.088$\pm$0.066 & 3.175$\pm$0.073 & 3.125$\pm$0.050\\
Flux (20-100 keV)\tablenotemark{a} & 4.76$\pm$0.25 & 4.41$\pm$0.25 & 4.60$\pm$0.18\\
$^{44}$Ti Line Flux\tablenotemark{b} & $-$1.65$\pm$1.88 & 7.09$\pm$4.20 & 1.57$\pm$2.81 \\
90\% Confidence Limit\tablenotemark{b}  & $\leq$1.85 & 0.72--8.99 & $\leq$3.59 \\
$\chi^2$/DOF & 68.0/51 & 34.0/51 & 60.8/51 \\
\enddata
\tablenotetext{a}{10$^{-11}$ ergs cm$^{-2}$s$^{-1}$}
\tablenotetext{b}{10$^{-5}$ photons cm$^{-2}$s$^{-1}$}
\end{deluxetable}

\clearpage


\figcaption[best_fit_with_lines.ps] {The top panel shows the counts
data with 1$\sigma$ error bars
for the combined HEXTE observations of Cas A. Data from both HEXTE clusters
have been combined and the data is binned as described in the body of this
paper. The best fit model is given by the solid line. The bottom panel gives
the $\chi$ value of the fit to each bin. \label{fig:fit}}

\clearpage

\begin{figure}
\epsscale{0.8}
\plotone{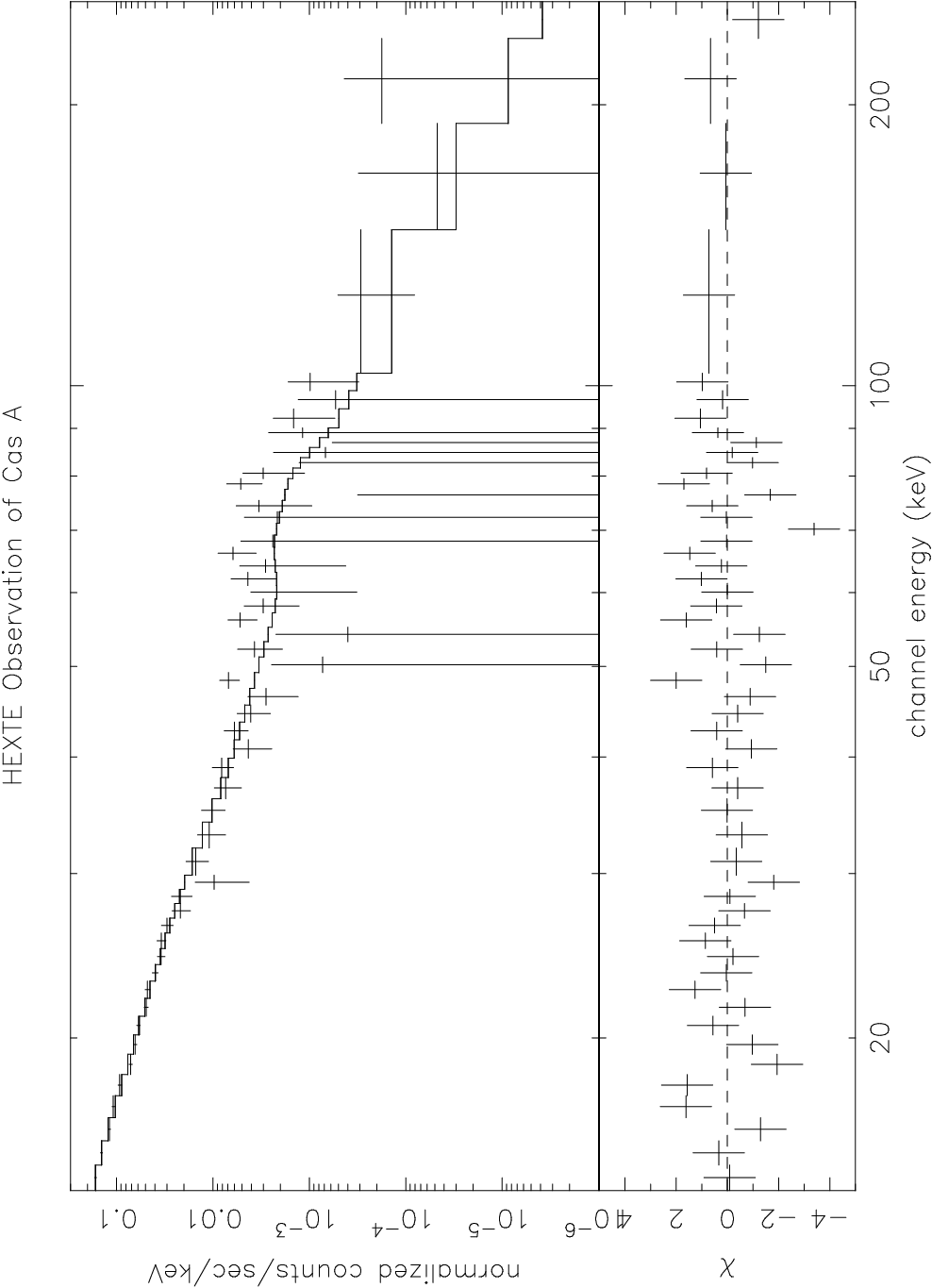}
\end{figure}

\end{document}